\documentclass[1p]{elsarticle}
\usepackage{fullpage}
\usepackage{amsmath}
\usepackage{amsfonts}
\usepackage{amssymb}
\usepackage{hyperref}
\usepackage{enumerate}

\newcommand{\IGNORE}[1]{}

\newcommand{\opt}{\mbox{\textsc{opt}\/}}

\begin{document}

{
\begin{frontmatter}
{
\title{A simple proof of the Moore-Hodgson Algorithm for minimizing the number of late jobs}

\author[uw]{Joseph Cheriyan\corref{cor1}\fnref{fn1}
}
\ead{jcheriyan@uwaterloo.ca}

\author[cmu]{R.~Ravi\fnref{fn2}
}
\ead{ravi@cmu.edu}

\author[tub]{Martin Skutella\fnref{fn3}
}
\ead{martin.skutella@tu-berlin.de}

\cortext[cor1]{Corresponding author}

\fntext[fn1]{This author acknowledges support from the Natural Sciences
\& Engineering Research Council of Canada (NSERC), No.~RGPIN-2019-04197.}

\fntext[fn2]{This material is based upon work supported in part by the U. S. Office of Naval
Research under award number N00014-21-1-2243 and the Air Force Office of
Scientific Research under award number FA9550-20-1-0080.}

\fntext[fn3]{Supported by the Deutsche Forschungsgemeinschaft (DFG, German Research Foundation) under Germany’s Excellence Strategy -- The Berlin Mathematics Research Center MATH+ (EXC-2046/1, project ID: 390685689).}

\address[uw]{
C\&O Dept., University of Waterloo, 
Waterloo, ON, Canada N2L~3G1
}

\address[cmu]{
Tepper School of Business,
Carnegie Mellon University, Pittsburgh, USA
}

\address[tub]{
Institute of Mathematics, Technische Universit\"{a}t Berlin, Germany
}

}


\begin{abstract}
{
The Moore-Hodgson Algorithm minimizes the number of late jobs on a single machine.
That is, it finds an optimal schedule for the classical problem $1~|\;|~\sum{U_j}$.
Several proofs of the correctness of this algorithm have been published.
We present a new short proof.
}
\end{abstract}

\begin{keyword}
Scheduling theory \sep Moore-Hodgson Algorithm \sep number of late jobs
\end{keyword}

\end{frontmatter}
}


\newtheorem{theorem}{Theorem}
\newtheorem{lemma}[theorem]{Lemma}
\newtheorem{propos}[theorem]{Proposition}
\newtheorem{corollary}[theorem]{Corollary}
\newtheorem{fact}[theorem]{Fact}
\newtheorem{claim}[theorem]{Claim}

\newenvironment{proof}{{\noindent \bf Proof:}}{\hfill$\Box$}

\newenvironment{remark}{{\noindent \bf Remark}:}{}



\section{Introduction} \label{sec:intro}

In 1968, J.~M.~Moore \cite{moore} presented an algorithm and analysis for
minimizing the number of late jobs on a single machine.
Moore stated
``The algorithm developed in this paper, however, consists of only
two sorting operations performed on the total set of jobs, \dots{}
%
%
Consequently, this method will be computationally feasible for very
large problems and can be performed manually on many smaller
problems.''
At the end of the paper, Moore presented a version of his algorithm
that he attributed to T.~E.~Hodgson; we follow that version.
In hindsight, the algorithm is ``just right'' for the problem, and
it is a popular topic in courses on Scheduling.
Several proofs of correctness have been published in the literature,
see, e.g., \cite{moore,french,handbook,pinedo}.
But, in our opinion, none of these proofs matches the simplicity
of the algorithm.  We present a proof that, hopefully, remedies
this discrepancy.

Our notation usually follows the notation of Pinedo \cite{pinedo}.
For a positive integer $\ell$, we use $[\ell]$ to denote the set $\{1,2,\dots,\ell\}$.

An instance $I$ of the scheduling problem $1~|\;|~\sum{U_j}$ consists of one machine
and $n$ jobs; the jobs are denoted $1,\dots,n$ (we identify a job
with its index).  Each job $j$ has a non-negative processing time $p_j$ and a
non-negative due date~$d_j$.

A schedule for this problem is a permutation of the $n$ jobs.  For
a given schedule $S$, the completion time of job $j$, denoted $C_j$,
is the sum of the processing times of job $j$ and the processing
times of the jobs that precede $j$ in $S$.  A job $j$ is called
\textit{late} (in the schedule $S$) if $C_j > d_j$.
\IGNORE{
, and the \textit{lateness}
of a job is defined to be $L_j=C_j-d_j$
}
The goal is to find a
schedule such that the number of late jobs is minimum.
We use $\opt(I)$ or $\opt$ to denote the minimum number of late
jobs of the instance $I$ (over all possible schedules).

A key feature of our proof is that we do not use induction on a
particular instance (which is the plan of Moore's proof), and
instead, we use induction on $\opt$ over all instances.

\section{The algorithm and analyis} \label{sec:algo}

The \textit{EDD~rule} (earliest due date rule) orders the jobs in
non-decreasing order of their due dates; this results in an
\textit{EDD~sequence}.  From here on, we assume that the jobs are
indexed according to the EDD~rule; that is,~$d_1\leq d_2\leq \dots\leq d_n$.

{
\begin{propos}\label{prop1}
If the EDD~sequence has a late job, then $\opt$ is $\geq1$.
\end{propos}

\begin{proof}
Let $k$ be the first late job in the EDD~sequence.
Thus, $C_k=\sum_{i\in[k]}p_i>d_k=\max_{i\in[k]}d_i$.
Consider any schedule $S$.
Let $\ell$ be the last of the jobs in $[k]$ in $S$.
Then, the completion time of $\ell$ in $S$ is at least $\sum_{i\in[k]}p_i>d_{\ell}$.
Thus $\ell$ is a late job of $S$.
\end{proof}
}

Clearly, if there is a sub-instance $I'$ that consists of a subset of the jobs
such that $\opt(I')=0$, then, the EDD~sequence of $I'$ has no late jobs.

The Moore-Hodgson Algorithm applies a number of iterations.
Each iteration maintains an EDD~sequence $\sigma$ of a subset of the jobs.
Initially, $\sigma=1,2,\dots,n$.
Each iteration either rejects one job from the sequence $\sigma$,
or terminates with the guarantee that $\sigma$ has no late jobs.
The algorithm finishes by outputting the concatenated schedule $\sigma, \zeta$,
where $\sigma$ is the sequence from the last iteration (that has no late jobs),
and $\zeta$ is an arbitrary permutation of all the rejected (i.e., late) jobs.

At the start of each iteration, $\sigma$ is an EDD~sequence of the non-rejected jobs.
An iteration of the algorithm examines the sequence of jobs
$\sigma_1,\sigma_2,\dots,\sigma_{\ell}$ (where $\ell\leq{n}$),
and finds the smallest index $k$ such that the job $\sigma_k$ is late
(thus, $C_{\sigma_k}>d_{\sigma_k}$ and $C_{\sigma_j}\leq{d_{\sigma_j}},\;\forall{j\in[k-1]}$).
The iteration terminates if there are no late jobs; otherwise, it
examines the ``prefix'' subsequence $\sigma_1,\dots,\sigma_k$, 
picks an index $m$ such that $p_{\sigma_m}$ is maximum among
$p_{\sigma_1},\dots,p_{\sigma_k}$, and rejects the job $\sigma_m$.

The following example illustrates the working of the algorithm; the
example is from Moore's paper~\cite{moore}.
There is one ``Completion time'' row for each iteration.
Whenever a job is rejected, its index is noted in the right-most
column, and its completion time in subsequent iterations is indicated
by an asterisk.
\begin{center}
\begin{tabular}{|rrrrrrrrr|l|}
\hline
EDD sequence:		&  1 &  2 &  3 &  4 &  5 &  6 &  7 &  8 & Rejected~Jobs \\
Due date $d_j$:		&  6 &  8 &  9 & 11 & 20 & 25 & 28 & 35 & \\
Processing time $p_j$:	&  4 &  1 &  6 &  3 &  6 &  8 &  7 & 10 & \\
\hline
Completion time $C_j$:	&  4 &  5 & 11 & &&&&& \\
                $C_j$:	&  4 &  5 & $\ast$ & &&&&& 3 \\
                $C_j$:	&  4 &  5 & $\ast$ & 8 & 14 & 22 & 29 && 3 \\
                $C_j$:	&  4 &  5 & $\ast$ & 8 & 14 & $\ast$ & 21 && 3, 6 \\
                $C_j$:	&  4 &  5 & $\ast$ & 8 & 14 & $\ast$ & 21 & 31 & 3, 6 \\
\hline
\end{tabular}	
\end{center}

\begin{theorem}\label{thm:correctness}
The Moore-Hodgson Algorithm outputs an optimal schedule for the problem $1~|\;|~\sum{U_j}$.
\end{theorem}

Our proof of Theorem~\ref{thm:correctness} is based on the following result.

\begin{lemma}\label{lem1}
Assume that there are late jobs in the EDD sequence~$\sigma=1,2,\dots,n$.
Let~$k$ be the first late job, and let~$m\in[k]$ be the job rejected
by the Moore-Hodgson Algorithm, i.e.,~$p_m=\max_{i\in[k]}p_i$.
There is an optimal schedule~$\pi$ that rejects job~$m$.

\end{lemma}

\begin{proof}
Consider an optimal schedule~$\pi$. 
Let~$R_\pi\subseteq[n]$ denote its subset of rejected (i.e., late)
jobs, and let $A_\pi:=[n]\setminus R_\pi$ denote its subset of on-time
(i.e., non-late) jobs.
By Proposition~\ref{prop1}, we may assume that $\pi$ schedules the
jobs in~$A_\pi$ in EDD order first, followed by the jobs in~$R_\pi$
in arbitrary order.

If~$m\in R_\pi$, we are done.

Otherwise, by Proposition~\ref{prop1}, there is a job~$r\in
R_\pi\cap([k]\setminus\{m\})$.  Consider the schedule~$\pi'$ that
sequences the jobs in~$A_{\pi'}:=(A_\pi\setminus\{m\})\cup\{r\}$
in EDD order first, followed by the jobs in~$[n]\setminus
A_{\pi'}=(R_\pi\setminus\{r\})\cup\{m\}$ in arbitrary order.
We will prove that~$\pi'$ schedules all jobs in~$A_{\pi'}$ on time.
It is thus optimal since~$|R_{\pi'}|=|R_\pi|$; moreover, by
construction,~$\pi'$ rejects job~$m$.

First, the jobs in~$A_{\pi'}\cap[k-1]$ are completed on time since
the EDD rule completes all jobs in~$[k-1]$ on time. Second, if~$k\in
A_{\pi'}$, then its completion time is
$\leq\sum_{i\in[k]\setminus\{m\}}p_i\leq\sum_{i\in[k-1]}p_i\leq
d_{k-1}\leq{d_k}$.
Third, compared to the former schedule~$\pi$, the completion
times of jobs in~$A_{\pi'}\setminus[k]=A_\pi\setminus[k]$ have been
changed in the new schedule~$\pi'$ by~$p_r-p_m\leq0$, hence,
these jobs also remain on time.
\end{proof}

\bigskip
\begin{proof} [of Theorem~\ref{thm:correctness}]
We use induction on $\opt$ over all instances of the problem.

\noindent
Induction basis:
If an instance has $\opt=0$, then by Proposition~\ref{prop1},
the algorithm outputs the EDD sequence with no late jobs.

\noindent
Induction step:
Let $I$ be an instance with $\opt(I)\geq1$ late jobs, and let
$I^{\ominus}$ be obtained from $I$ by deleting the job~$m$ rejected
in the first iteration of the algorithm.
By Lemma~\ref{lem1},~$\opt(I^{\ominus})\leq\opt(I)-1$.
Thus, by the induction hypothesis, the algorithm finds an optimal
schedule $S^{\ominus}$ for $I^{\ominus}$.  The algorithm for $I$
outputs the schedule $S$ such that $S$ is the same as $S^{\ominus}$
except that job $m$ is added at the end, as a rejected job.  Clearly,
$S$ has at most $\opt(I^{\ominus})+1\leq\opt(I)$ rejected jobs and
is thus optimal.
\end{proof}

\bigskip
\begin{remark}
Some of the previous proofs (\cite{moore}, \cite{french},
\cite[2nd~edition]{pinedo}, \cite{maxwell}, \cite{sturm}) use an induction-type argument on
a particular instance; then, the argument has to track the parameters
of the sequence of rejected jobs throughout; this is not difficult,
but, a rigorous presentation takes more than one page.

\end{remark}

\begin{remark}
Our proof generalizes to the problem of $1~|\;|~\sum{w_jU_j}$ where jobs $j$ are provided with a weight $w_j$ in addition to their processing time and the minimization objective is now the weighted number of late jobs, in the special case when the the processing times and job weights are oppositely ordered: i.e., $p_i \leq p_j$ implies $w_i \geq w_j$ (see \cite{elements}).
The key observation is that in the proof of Lemma~\ref{lem1} when we replace the alternate choice $r$ that was rejected  in $R_{\pi}$ by the correct choice $m$ in the first bad prefix $[k]$, the opposite ordering relation implies that since $p_r \leq p_m$, then $w_r \geq w_m$. Since we replace $w_r$ in the weighted objective with the potentially smaller $w_m$, this change also makes the weighted objective no worse. The remaining elements of the proof are unchanged.
\end{remark}


\bigskip
\bigskip
\noindent
{\bf Acknowledgments}:
We thank several colleagues for insightful discussions over the
years. The idea of the presented proof stems from a discussion
between the authors at the Tenth Carg\`ese Workshop on Combinatorial
Optimization at the Institut d'\'Etudes Scientifiques de Carg\`ese,
Corsica (France). We are much indebted to the organizers of the
workshop, in particular for putting the focus of the 2019 edition
on ``Proofs from the Book in Combinatorial Optimization.''



\end{document}